\documentclass[prd,twocolumn,showpacs]{revtex4}
\usepackage{hyperref,amssymb,amsmath,mathrsfs,bm,graphicx}
\usepackage{enumitem}
\usepackage{color}
\begin{document}
\title{Physical properties of a source of the Kerr metric: Bound on the surface gravitational  potential  and conditions for the fragmentation} 
\author{L. Herrera}
\email{lherrera@usal.es}
\affiliation{Instituto Universitario de F\'isica
Fundamental y Matem\'aticas, Universidad de Salamanca 37007, Salamanca, Spain and Escuela de F\'\i sica, Facultad de Ciencias, Universidad Central de Venezuela, Caracas 1050, Venezuela}
\author{J. L. Hernandez-Pastora}
\email{jlhp@usal.es}
\affiliation{Departamento de Matem\'atica Aplicada and Instituto Universitario de
F\'\i sica Fundamental y Matem\'aticas, Universidad de Salamanca, Salamanca, Spain}

\begin{abstract}
We investigate some important physical aspects of a recently presented interior solution for the Kerr metric. It is shown that, as in the spherically symmetric case, there is a specific limit for the maximal value of the surface potential (degree of compactness), beyond which, unacceptable  physical anomalies appear. Such a bound is related to the appearance of negative (repulsive)  gravitational acceleration, that is accompanied by the appearance of negative values of the pressure. A detailed discussion on this effect is presented. We also study the  possibility of a fragmentation scenario, assuming that the source leaves the equilibrium, and we bring out the differences   with the spherically symmetric case.
\end{abstract}
\date{\today}
\maketitle

\section{Introduction}
In a recent paper \cite{kerrinterior}, we have proposed an interior solution for the Kerr metric \cite{Kerr}, satisfying the matching conditions on the boundary surface of the matter distribution, and endowed with reasonable physical properties, at least for a range of values of the parameters, which includes  values considered in the existing literature to describe realistic models of rotating neutron stars and white dwarfs. 

 For these reasons, after decades of intense theoretical work  (see \cite{Her1, C, Wa, Her2, W, Kr, HM, HJ, H, DT, Pa, Via1, Via2, K, Az, K1, maj} and references therein), the solution presented in \cite{kerrinterior}, may be reasonably  regarded as a satisfactory solution to  the problem of constructing  a physically viable source for the Kerr metric.

It is the purpose of this work, to study some important physical properties of the source mentioned above. 

We shall first establish a bound on the degree of compactness (surface gravitational potential), which of course implies  a bound on the  gravitational surface redshift of spectral lines from the surface of our source, similar to the limit existing for spherically symmetric sources \cite{Buch, B1}. Two points deserve to be emphasized here:
\begin{itemize}
\item Our source is generated by an anisotropic fluid. 
\item In the spherically symmetric case, there is a well established link between the maximal values of the surface redshift, and the local anisotropy of pressure,   (see \cite{chew81, mx1, mx2,14,  mx3, mx5, mx6, mx4, mx7, 2p, lake} and references therein).The great interest aroused  on this issue is easily justified, if we recall that the surface redshift is an observable variable,  thereby entitled to provide relevant information about the structure of the source. 

\end{itemize}

In the spherically symmetric case, the bound on the  degree of compactness expresses itself  through the appearance of different kind of physical anomalies which render the fluid distribution physically inviable (e.g. singularities of the physical variables, negative energy density, etc.).

In this manuscript, we shall relate  the above mentioned limit,  to the appearance of  repulsive gravitational acceleration, described by means of the acceleration tensor recently introduced by Maluf \cite{maluf}.  This  tensor gives the values of the inertial (i.e., non-gravitational) accelerations that are necessary to maintain the frame in a given inertial state (stationary in our case). If the frame is maintained stationary in space-time, then the inertial acceleration is exactly minus the gravitational acceleration imparted to the frame. Obviously, once the acceleration tensor becomes negative, for any piece of the material, the system becomes unstable, and leaves the stationary state.

We shall investigate in detail what is the maximal  degree of compactness which excludes the appearance of negative gravitational acceleration, for a range of values of the angular momentum of the source. We shall also see that the appearance of negative gravitational accelerations is always accompanied  by the appearance of negative pressure of the fluid distribution and a singularity of the pressure at the center of the fluid distribution, for exactly the same values of the parameters.

Next, we shall consider our source as the initial state of a fluid distribution which is assumed to leave the equilibrium regime. We shall then evaluate  the system on a time scale that  is smaller than the hydrostatic time scale. Doing so we shall be able to detect possible scenarios  of fragmentation (cracking) of the source, which are absent in the spherically symmetric limit of our source.

\section{A source for the exterior Kerr solution}
In \cite{kerrinterior} we provided a general method to construct global models of self--gravitating stationary sources.  As a particular example we considered a source for the Kerr metric. The present study concerns this last  fluid distribution. In what follows we shall very briefly summarize the basic equations and the main properties  of the solution. We refer the reader to \cite{kerrinterior} for the details and some intermediate calculations. At  this point we would like to call  attention to two misprints in \cite{kerrinterior}, that have been corrected here, namely: the equation (\ref{r12}) below,  is the right version of  the corresponding equation (36) appearing in \cite{kerrinterior}. Also,  a misprint appearing in the  equation (40) in \cite{kerrinterior}, has been corrected in the last of  the forthcoming equations (\ref{prepsimq1}). It is worth emphasizing, however, that such  misprints are irrelevant for the discussion presented in \cite{kerrinterior}, since all the calculations in that reference were carried out using the right expressions, written down here.

\subsection{The exterior metric}

The  line element for a vacuum stationary  and axially symmetric space--time, in Weyl canonical coordinates, may be written as :
\begin{equation}
 ds^2_E=-e^{2\psi}(dt-w d\phi)^2+e^{-2\psi+2\Gamma}(d\rho^2+dz^2)+e^{-2\psi}\rho^2d\phi^2,
\label{1}
\end{equation}
where $\psi=\psi(\rho,z)$ ,  $\Gamma=\Gamma(\rho,z)$ and  $w=w(\rho,z)$ are functions of their arguments.

For vacuum space--times, Einstein's field equations imply for the
metric functions 
\begin{equation}f( f_{, \rho \rho}+\rho^{-1}
f_{, \rho}+f_{, zz})-f_{, \rho}^2-f_{, z}^2+\rho^{-2} f^4(w_{, \rho}^2+w_{, z}^2) = 0, \label{meq1}
\end{equation}
\begin{equation}f( w_{, \rho \rho}+\rho^{-1}
w_{, \rho}+w_{, zz})+2f_{, \rho}w_{, \rho}+2f_{, z}w_{, z} = 0, \label{meq2}
\end{equation}
with $f \equiv e^{2\psi}$ and
\begin{eqnarray}
\Gamma_{, \rho}&=&\frac 14 \rho f^{-2} (f_{, \rho}^2-f_{, z}^2)-\frac 14 \rho^{-1} f^{2} (w_{, \rho}^2-w_{, z}^2) \nonumber \\ 
\Gamma_{, z}&=& \frac 12 \rho f^{-2} f_{, \rho}f_{, z}-\frac 12 \rho^{-1} f^{2} w_{, \rho}w_{, z}. \label{meq3}
\end{eqnarray}

It will be useful to introduce the   Erez-Rosen \cite{erroz}, or standard Schwarzschild--type coordinates $\{r,y\equiv \cos\theta\}$ or in  spheroidal prolate coordinates $\{x\equiv\frac{r-M}{M},y\}$ \cite{quev}:
\begin{equation}
\rho^2=r(r-2M)(1-y^2) \ , \quad z=(r-M)y,
\label{2}
\end{equation}
where $M$ is a constant which will be identified later.

In terms of the above coordinates the line element (\ref{1}) may be written as:
\begin{widetext}
\begin{eqnarray}
 ds^2_E=-e^{2\psi(r,y)}(dt -w d\phi)^2+e^{-2\psi+2\left[\Gamma(r,y)-\Gamma^s
 \right]}dr^2+
 e^{-2\left[\psi-\psi^s\right]+
 2\left[\Gamma(r,y)-\Gamma^s\right]}r^2 d\theta^2
+ e^{-2\left[\psi-\psi^s\right]} r^2\sin^2\theta d\phi^2,
\label{exterior}
\end{eqnarray}
\end{widetext}

where  $\psi^s$ and $\Gamma^s$ are the metric functions corresponding to the Schwarzschild solution, namely,
\begin{equation}
\psi^s=\frac 12 \ln \left(\frac{r-2M}{r}\right) \, \quad \Gamma^s=-\frac 12 \ln \left[\frac{(r-M)^2-y^2M^2}{r(r-2M)}\right],
\label{schwfun}
\end{equation}
where the parameter $M$ is easily identified as the Schwarzschild mass.

For the specific case of the Kerr metric, in Weyl  coordinates,  we have  the following expressions for the metric functions
\begin{equation}
f=\frac{(r_1+r_2)^2(1-j^2)-4 M^2(1-j^2)+j^2(r_1-r_2)^2}{(r_1+r_2+2M)^2(1-j^2)+j^2(r_1-r_2)^2},
\label{fkerr}
\end{equation}
\begin{equation}
e^{2\Gamma}=\frac{(r_1+r_2)^2(1-j^2)-4 M^2(1-j^2)+j^2(r_1-r_2)^2}{4r_1r_2(1-j^2)},
\label{gakerr}
\end{equation}
\begin{equation}
w=\frac{j(2M+r_1+r_2)(4M^2(1-j^2)-(r_1-r_2)^2)}{(r_1+r_2)^2(1-j^2)-4 M^2(1-j^2)+j^2(r_1-r_2)^2},
\label{omkerr}
\end{equation}
where $j\equiv \frac{J}{M^2}=a/M$ denotes the dimensionless parameter representing the angular momentum of the source, and  is related  to  the rotation parameter $a$ of the Kerr metric in its well-known Boyer-Lindquist representation.

 Also, $r_{1,2}\equiv \displaystyle{\sqrt{\rho^2+
(z\pm M\sqrt{1-j^2})^2}}$, which in the Erez-Rosen coordinates become
\begin{equation}
r_{1,2}^2=\left[(r-M)\pm My\sqrt{1-j^2}\right]^2-M^2 j^2(1-y^2).
\label{r12}
\end{equation}

 From the study of the relativistic multipole moments (RMM) \cite{geroch, ger, ger3, han, th, fhp},  of the Kerr solution, it follows that 
\begin{equation}
m_k=M_k=M (i a)^k,
\end{equation}
where   ($m_k$) are the   expansion coefficients of the Ernst potential on the axis of symmetry. 

Also, we have that, the massive RMM (even orders)  and the rotational RMM (odd orders) can be expressed as:
\begin{equation}
M_{2l}=(-1)^l M^{2l+1} j^{2l} \quad , \quad M_{2l+1}=i (-1)^l M^{2l+2} j^{2l+1}.
\label{RMMkerr}
\end{equation}

From the above it follows  that the rotation of the object leads to a negative quadrupole massive moment, $ {\displaystyle q\equiv \frac{M_2}{M^3}=- j^2}$, i.e. all the possible sources of Kerr solution are oblate (see \cite{kerrinterior}  for details).

\subsection{The interior metric}
Following the procedure sketched in \cite{kerrinterior}, the following line element is assumed at the interior:
\begin{eqnarray}
ds^2_I&=&-e^{2 \hat a} Z(r)^2 ( dt -\Omega d\phi)^2+\frac{e^{2\hat g-2\hat a}}{A(r)} dr^2+e^{2\hat g-2\hat a}r^2 d\theta^2\nonumber \\&+&e^{-2 \hat a}r^2 \sin^2\theta d\phi^2,
\label{interior}
\end{eqnarray}
with
\begin{equation}
\hat a \equiv a(r,\theta)-a^s(r) \ , \qquad  \hat g \equiv g(r,\theta)-g^s(r,\theta),
\label{funcint}
\end{equation}
 where $\Omega=\Omega(r,\theta)$, and $a^s(r)$ and  $g^s(r,\theta)$ are functions that, on the boundary surface, equal the metric functions corresponding to the Schwarzschild solution  (\ref{schwfun}), i.e.  $a^s(r_{\Sigma})=\psi^s_{\Sigma}$ and  $g^s(r_{\Sigma})=\Gamma^s_{\Sigma}$.  Also, $A(r)\equiv 1-pr^2$  and  $Z\equiv\displaystyle{ \frac 32 \sqrt{A(r_{\Sigma})}-\frac 12 \sqrt{ A(r)}}$, where $p$ is an arbitrary constant and the boundary surface of the source is defined by $r=r_{\Sigma}=const.$ 
 
The case $w=0$, $\hat g=\hat a=0$, corresponds to a spherically symmetric distribution, more specifically, to the well known incompressible (homogeneous energy density) perfect fluid sphere, and hence the matching of (\ref{interior}) with the Schwarzschild solution implies $p=\displaystyle{\frac{2M}{r_{\Sigma}^3}}$. The simple condition $w=0$ recovers, of course, the static case.

It should be noticed  that for  simplicity we consider here only matching surfaces of the form $r = r_\Sigma =
const$. In particular, this choice simplifies considerably the treatment of the matching conditions (see below).  Besides, this  type of boundary surface allows to describe quite appropriately the shape, that we expect for a relativistic rotating body.  Of course more general surfaces with axial symmetry, of the form $r = r_\Sigma(\theta)$, could be considered as well.

In order to satisfy  the matching (Darmois) conditions \cite{26},  the following equations have to be satisfied: 
\begin{eqnarray}
&& a_{\Sigma}=\psi_{\Sigma} \ , \quad a^{\prime}_{\Sigma}=\psi^{\prime}_{\Sigma} \ , \quad g_{\Sigma}=\Gamma_{\Sigma} \ , \quad g^{\prime}_{\Sigma}=\Gamma^{\prime}_{\Sigma}, \nonumber\\ 
&&a^s_{\Sigma}=\psi^s_{\Sigma} \ , \quad (a^s)^{\prime}_{\Sigma}=(\psi^s)^{\prime}_{\Sigma}, \nonumber \\ && g^s_{\Sigma}=\Gamma^s_{\Sigma} \ , \quad (g^s)^{\prime}_{\Sigma}=(\Gamma^s)^{\prime}_{\Sigma},
 \nonumber\\
 && \Omega_{\Sigma}=w_{\Sigma} \ , \quad \Omega^{\prime}_{\Sigma}=w^{\prime}_{\Sigma},
\label{matchingcond}
\end{eqnarray}
where prime denote partial derivative with respect to $r$  and subscript $\Sigma$ indicates that the quantity is evaluated on the boundary surface. It is important to keep in mind that we are using global coordinates $\{r,\theta\}$  on both sides of the boundary.

Indeed, the Darmois matching conditions require the continuity of the first and the second fundamental form across the boundary surface of the source. 

	The first fundamental form is just the induced metric on the boundary surface. Therefore the first set of Darmois conditions requires:
\begin{equation}
(ds^2_E)_\Sigma-(ds^2_I)_\Sigma\equiv[ds^2]\stackrel{\Sigma}{=}0,
\label{fs}
\end{equation}
where $ds^2_E$ and $ds^2_I$ are given by (\ref{exterior}) and (\ref{interior}), respectively, and the square bracket denotes the discontinuity of any enclosed quantity, across the boundary surface of the source.
Now, since the matching surface considered is  $r = r_\Sigma =const$, the first fundamental form is continuous on that surface  whenever $\displaystyle { a_{\Sigma}=\psi_{\Sigma}}$, $\displaystyle{ g_{\Sigma}=\Gamma_{\Sigma}}$ and $\displaystyle{\Omega_{\Sigma}=w_{\Sigma}}$. 

Next, we have to  require the continuity of the second fundamental form, which implies that the extrinsic curvature evaluated  on both sides of the matching surface must be equal.

This second fundamental form ($II$) evaluated on the boundary surface, is defined by 
\begin{equation}
II=-(n_{\mu;\nu}dx^{\nu}dx^{\mu})_\Sigma \equiv (K_{ab}dx^a dx^b)_\Sigma,
\label{ss}
\end{equation}
where  the indexes $a$,$b$ stand for $\{t,\theta,\phi\}$, and $n_\mu$ denotes the unit, normal vector to the boundary surface.

In our case, the boundary surface equation is given by
\begin{equation}
f\equiv r-r_\Sigma=0; \qquad  r_\Sigma=constant,
\label{nv}
\end{equation}
implying that the unit vector, normal to the boundary surface is defined by:
\begin{equation}
n_\mu=\frac{\partial_\mu f}{\sqrt{\partial_\alpha f \partial_\beta f g^{\alpha \beta}}}.
\label{nv3}
\end{equation}

  From (\ref{nv}) and (\ref{nv3}) it follows that  $K_{ab} =-\Upsilon^1_{ab}$,  where $\Upsilon^i_{jk}$, denote  the Christoffel symbols of the corresponding (interior or exterior) metric.  Then, after some simple calculations we obtain:
	\begin{widetext}
\begin{eqnarray}
0\stackrel{\Sigma}{=}\left[K_{tt}\right]&\Rightarrow& e^{4\hat a-2\hat g}A\left(\hat a^{\prime} Z^2+ZZ^{\prime}\right)\stackrel{\Sigma}{=}e^{4\psi-2\hat \Gamma}\left(\psi^{\prime}\right) , \nonumber\\ 
0\stackrel{\Sigma}{=}\left[K_{\theta \theta}\right]&\Rightarrow&A\left((\hat g^{\prime}-\hat a^{\prime})r^2+r \right)\stackrel{\Sigma}{=}e^{2\psi^s}\left((\hat {\Gamma}^{\prime}-\hat{\psi}^{\prime})r^2+r\right) , \nonumber \\ 
0\stackrel{\Sigma}{=}\left[K_{\phi \phi}\right]&\Rightarrow& (-\hat{\psi}^{\prime}r^2+r) \sin^2\theta \  e^{-2\hat{\psi}}-(\psi^{\prime}w^2+w w^{\prime})e^{-2\psi}\stackrel{\Sigma}{=}\nonumber \\
&&(-\hat{a}^{\prime}r^2+r) \sin^2\theta \ e^{-2\hat{a}}-(\hat a^{\prime}Z^2\Omega^2+Z Z^{\prime}\Omega^2+Z^2\Omega \Omega^{\prime})e^{2\hat a},
\nonumber\\
0\stackrel{\Sigma}{=}\left[K_{\phi t}\right]&\Rightarrow&2\psi^{\prime}w+w^{\prime}\stackrel{\Sigma}{=}
(2\hat a^{\prime}Z^2\Omega+2Z Z^{\prime}\Omega+Z^2 \Omega^{\prime})e^{-2\psi^s} ,
\label{secondFF}
\end{eqnarray}
\end{widetext}
where $\hat{\psi}\equiv \psi-\psi^s$, $\hat{\Gamma}\equiv \Gamma-\Gamma^s$ and the square bracket denotes the discontinuity of any enclosed quantity, across the boundary surface of the source.
From the above expressions, it follows at once that the continuity of $K_{tt}$ requires  that $ \hat a^{\prime}_{\Sigma}=\hat{\psi}^{\prime}_{\Sigma}$. The continuity of $K_{\theta \theta}$  implies  that $ \hat g^{\prime}_{\Sigma}=\hat{\Gamma}^{\prime}_{\Sigma}$, and finally the continuity of 	$K_{\phi \phi}$ and $K_{\phi t}$ imposses that $\Omega^{\prime}_{\Sigma}=w^{\prime}_{\Sigma}$. These conditions, together with those obtained from the continuity of the first fundamental form,  are exactly conditions (\ref{matchingcond}).

In the  spherically symmetric case, we have  $\hat a=\hat g=0$, and the physical variables are obtained from the field equations for a perfect fluid, the result is well known and reads (in relativistic units)
\begin{eqnarray}
-T^0_0\equiv \mu&=&\frac{3p}{8 \pi},\nonumber\\
T^1_1=T^2_2=T^3_3\equiv P&=&\mu\left(\frac{\sqrt A-\sqrt{A_{\Sigma}}}{3 \sqrt{A_{\Sigma}}-\sqrt{A}}\right),
\label{eeesf}
\end{eqnarray}
with  $A=\displaystyle{1-\frac{2m(r)}{r}=1-p r^2=1-\frac{2M r^2}{r_{\Sigma}^3}}$,
where $\mu$ and $P$ denote the energy density and the isotropic pressure respectively, and  for the mass function $m(r)$ we have
\begin{equation}
m(r)=-4\pi\int^{r}_0r^2 T^0_0 dr,
\end{equation}
implying 
\begin{equation}
M\equiv m(r_{\Sigma})=-4\pi\int^{r_{\Sigma}}_0r^2 T^0_0 dr=\frac{p r_{\Sigma}^3}{2}.
\end{equation}

This model, which describes the well known incompressible perfect fluid sphere, is further restricted by the requirement that the pressure be regular and positive everywhere within the fluid distribution, which implies $\displaystyle{\tau\equiv{\frac{r_{\Sigma}}{M}}>\frac94}$, where $\tau$ measures the inverse of the degree of compactness. As it is evident from (\ref{eeesf}) the pressure vanishes at the boundary surface. 

 We shall now proceed to consider the general, non--spherical case.  In \cite{kerrinterior} we provided a general procedure to choose the interior metric functions $\hat a$, $\hat g$ and $\Omega$ producing physically meaningful  models. With this aim,  besides the fulfilment of the junction conditions (\ref{matchingcond}), we  required that all physical variables be regular within the fluid distribution and the energy density be positive.  Following this procedure, we have for the interior of the Kerr metric:
\begin{eqnarray}
\hat a(r,\theta)&=&\hat \psi_{\Sigma} s^2(3-2s)   +r_{\Sigma}\hat \psi^{\prime}_{\Sigma}s^2(s-1),\nonumber \\
\hat g(r,\theta)&=&\hat \Gamma_{\Sigma} s^3(4-3s)   +r_{\Sigma}\hat \Gamma^{\prime}_{\Sigma}s^3(s-1),
\nonumber \\
\Omega(r,\theta)&=&w_{\Sigma} s^4(5-4s)+r_{\Sigma}w_{\Sigma}^{\prime} s^4(s-1).
\label{aygyomsimple}
\end{eqnarray}
with $s\equiv r/r_{\Sigma} \in \left[0,1\right]$, and where 

\begin{widetext}
\begin{eqnarray}
&&\hat \psi_{\Sigma}\equiv \psi_{\Sigma}-\psi^s_{\Sigma}=\frac 12\ln\left\lbrace\frac{\tau}{\tau-2}\frac{N+ r_1^{\Sigma}r_2^{\Sigma} (2j^2-1)}{N+ r_1^{\Sigma}r_2^{\Sigma} (2j^2-1)-2(1-j^2)(r_1^{\Sigma}+r_2^{\Sigma}+2)} \right\rbrace,\nonumber\\
&&\hat \Gamma_{\Sigma}\equiv \Gamma_{\Sigma}-\Gamma^s_{\Sigma}=\frac 12\ln\left\lbrace\frac{(\tau-1)^2-y^2}{\tau(\tau-2)}\frac{N+ r_1^{\Sigma}r_2^{\Sigma} (2j^2-1)}{2 r_1^{\Sigma}r_2^{\Sigma}(j^2-1)} \right\rbrace, \nonumber\\
&&w_{\Sigma}=M j \frac{(N+ r_1^{\Sigma}r_2^{\Sigma}) (2+r_1^{\Sigma}+r_2^{\Sigma})}{-N+ r_1^{\Sigma}r_2^{\Sigma}(1-2j^2)},
\label{prepsimq1}
\end{eqnarray}
\end{widetext}
with 
\begin{equation}
r_{1,2}^{\Sigma}=\sqrt{\left(\tau -1\pm y\sqrt{1-j^2}\right)^2-j^2 (1-y^2)}.
\end{equation}
\begin{equation}
N\equiv -\tau (\tau-2)+(1-y^2)-j^2,
\end{equation}

The metric functions so  obtained, satisfy the junction conditions  (\ref{matchingcond}) and produce physical variables (equations (17)--(21) in \cite{kerrinterior}), which are regular within the fluid distribution.  Furthermore  the vanishing of  $\hat g$ on the axis of symmetry, as required by the regularity conditions,  necessary to ensure elementary flatness in the vicinity of  the axis of symmetry, and in particular at the center \cite{1n}--\cite{3n},
 is assured by the fact that $\hat \Gamma_{\Sigma}$ and $\hat \Gamma^{\prime}_{\Sigma}$ vanish on the axis of symmetry. 
Also, the good behaviour of the function $\Omega$ on the symmetry  axis is fulfilled since $w_{\Sigma}$ and $w_{\Sigma}^{\prime}$ vanish when $y=\pm 1$.

\section{Bound on the surface redshift and the acceleration tensor}
We shall now proceed to establish the limit on the degree of compactness of our source. For doing so we shall make use of the concept of the acceleration tensor introduced in \cite{maluf}.

The acceleration tensor gives the values of the
inertial (i.e., non-gravitational) accelerations that are necessary to maintain the frame adapted to a field of observers in space-time, in a given inertial state. If the frame is maintained static (stationary)  in space-time, then the inertial acceleration is exactly minus the gravitational acceleration imparted to the frame.

Thus, we may write
\begin{equation}
{\vec a}= \frac{1}{2 (-g^{00})^{3/2}}\left(\frac{\partial_1 g^{00}}{\sqrt{g_{11}}} {\vec r}+\frac{\partial_2 g^{00}}{\sqrt{g_{22}}} {\vec \theta}\right),
\label{acel}
\end{equation}
with ${\vec a}\equiv a_r {\vec r}+a_{\theta} {\vec \theta}$.

In our case the following expression verifies, 
\begin{equation}
g^{00}=-\frac{e^{-2 \hat a}}{Z^2}+e^{2 \hat a}\frac{\Omega^2}{r^2\sin^2\theta},
\label{g00}
\end{equation}

then the radial component of the acceleration (the only one we need for our discussion), reads
\begin{widetext}
\begin{equation}
a_r = \frac{e^{-\hat g(s,y)}\sqrt{A}  }{ r_{\Sigma}(-g^{00})^{3/2}}\left[\frac{e^{-\hat a(s,y)}}{Z(s)^3}  \left(Z(s) \partial_s\hat a(s,y)+\frac{s}{\sqrt{\tau (\tau-2 s^2)}} \right) +e^{3 \hat a(s,y)}\frac{{\hat \Omega}^2}{s^2 \sin^2\theta}\left( \partial_s \hat a+\frac{ \partial_s \hat \Omega}{\hat \Omega}-\frac{1}{s}\right)  \right],
\label{compo}
\end{equation}
\end{widetext}
where  $s\equiv r/r_{\Sigma} \in [0,1]$, and  $\hat \Omega\equiv \Omega(s,y)=\Omega(r,y)/r_{\Sigma}$.

As we have already mentioned, for the spherically symmetric incompressible fluid, we have a bound for the degree of compactness (surface gravitational potential) given by $\displaystyle{\tau\equiv{\frac{r_{\Sigma}}{M}}>\frac94}$. Usually, such a limit is related to the appearance of a singularity of the pressure at the center of the distribution.

We shall start our study by imposing only  the positivity of the energy density and the absence of singularities in the physical variables. Doing so, the evaluation of (\ref{compo})	 produces   the following results:
\begin{itemize}

\item The change of sign in  $a_r$ (from positive to negative), occurs always for a critical value of $\tau_c$,  ($\tau_c=9/4=2.25$). A specific example is depicted in figure 1, for a given value of the parameter $j$ and a given value of the angular coordinate $\theta$. The continuous line corresponds to a value of  $\tau>\tau_c$. As  is apparent from this figure, the range of values of the radial coordinate $s$, for which the acceleration is negative, depends on the values of   $\tau$. Figure 2  is a plot of the acceleration as function of  $s$ and  $\tau$ for a given value of $y$. Again the negative values of the acceleration are clearly exhibited

\item The behaviour of   $a_r$ depends, although very slightly, on the angular coordinate $y$. This is shown in the figure 3, for vey small values of $\tau$ (very compact objects), for  the value of the radial coordinate  $s=0.3$. Observe the smaller (in absolute value) values of   $a_r$, close to the axis of symmetry. This can also be appreciated in the figure 4, for a value of  $\tau=2.1$.

\item In all the examples analyzed, the positive energy condition ( P.E.D. ) is always satisfied ($-T_0^0>0$), at least for  $j<0.1$. This is clearly indicated in the Table I, where minimal values of   $\tau$ (maximal degree of compactness), compatible with  P.E.D., are given for different values of $j$. As  is apparent in this table, for values $j<0.1$ the minimal value of $\tau$ preserving the P.D.E., is smaller than  $\tau_c=2.25$, thereby indicating that negative accelerations are compatible with positive energy density.

\item Although, the value of $j$ affects the value of   $a_r$ when it is positive, the former neither affects the value of  $a_r$ when it is negative, nor does it  affect  the value  of $s$ for which the acceleration changes of sign. This is clearly shown in the figure 5. It is worth noting that this behaviour also holds for $j=0$, as it is apparent from the corresponding curve in  figure 5. The general behaviour is always the same: negative acceleration appears in the inner part of the source, whereas it is positive at the outer one, and there is no range of values  of $\tau$ and/or  $j$, for which this behaviour inverts.

\item The issue mentioned in the point above, suggests that the  change of sign in   $a_r$  and the existence of a critical value  $\tau_c$, should be related with restrictions already appearing in the spherically symmetric case. This is indeed the case. In fact, below  $\tau_c$, the radial pressure becomes negative in some regions of the source and becomes singular at the origin. Furthermore, the range of the radial coordinate for which the acceleration becomes negative ($s\in [0,s_c]$), where the value of  $s_c$ depends on $\tau$ but not on $j$, is exactly the same range for which the radial pressure becomes negative.  

\item The exact value of  $s_c$, can be determined by evaluating the radial pressure (expression (19) in \cite{kerrinterior}). There is however a much simpler way of doing that. Indeed, as it  can be seen from  (\ref{eeesf}),  the pressure  $P$ (in the spherically symmetric case) is always positive, unless the denominator in that expression becomes negative. On the other hand, the small corrections to the radial pressure introduced by  the non--sphericity in the source under consideration, do not affect the conclusion above, since they are smaller than    $P$ in their absolute value (see fig 3 in  (\cite{kerrinterior}). In other words the radial pressure is negative only if  $P<0$. 

\item In figure 6 we plot the denominator of (\ref{eeesf}) as a function of $s$, for different values of $\tau$. The roots of these curves define the range  $s \in [0,s_c]$ for which the acceleration and the radial pressure become negative. Finally, in table II, we show some values of $s_c$ for different  values of $\tau$.
\end{itemize}

	\begin{table}[htp]
		\caption{$\tau_{min}$ denotes the minimal value of the inverse compactness factor compatible with the positive energy density  condition  Positive Energy Density (P.E.D.) ($-T_0^0>0$).}
		\begin{center}
			P.E.D. ($-T_0^0>0$)
			
			\begin{tabular}{cc}
				\hline\hline
					$  j$ \hspace*{2cm}&$\tau_{min} $   \\  \hline
				0.1 \hspace*{2cm}&2.27  \\  \hline
					0.01 \hspace*{2cm}&2.07  \\  \hline
						0.005\hspace*{2cm} &2.05   \\  \hline
							0.001\hspace*{2cm}&2.03 
				 \\ 
				\hline\hline
			\end{tabular}
		\end{center}
		\label{data}
	\end{table}

\begin{figure}[h]
	\includegraphics[scale=0.31]{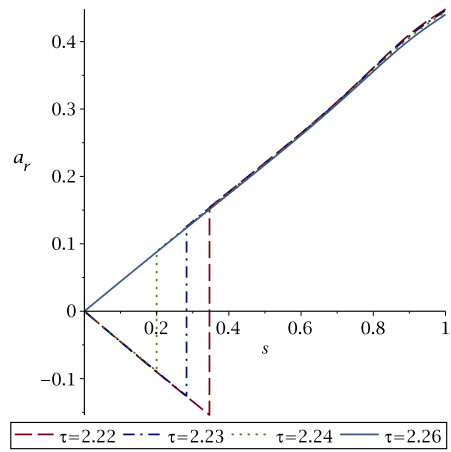} 
	\caption{\label{fig1} {\it   Graphics of  $a_r$ as function of $s$, for different values of  $\tau$,  with  $j=0.1$,   $y\equiv \cos\theta=0.5$.}}
\end{figure}

\begin{figure}[h]
	\includegraphics[scale=0.31]{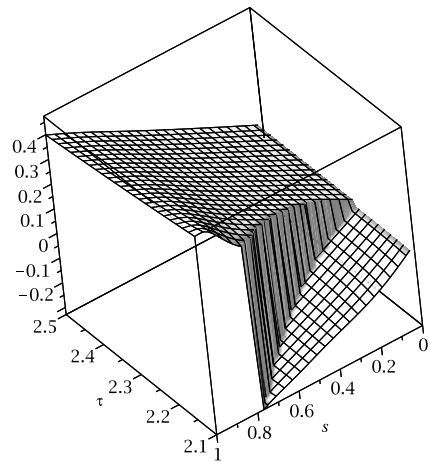} 
	\caption{\label{fig2} {\it Plot of $a_r$  as function of $s$ and $\tau$, for  $j=0.1$,  $y=0.2$.}}
\end{figure}

\begin{figure}[h]
	\includegraphics[scale=0.31]{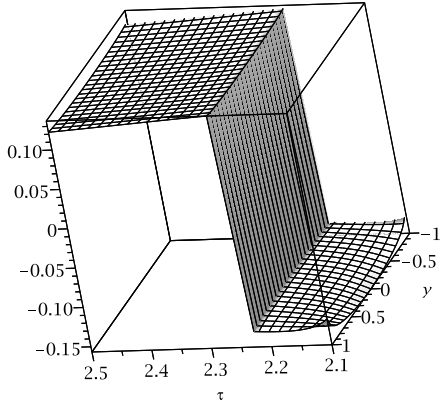} 
	\caption{\label{fig3} {\it Plot of  $a_r$  as function of  $\tau$ and  $y$, with $j=0.1$,    $s=0.3$.}}
\end{figure}

\begin{figure}[h]
	\includegraphics[scale=0.31]{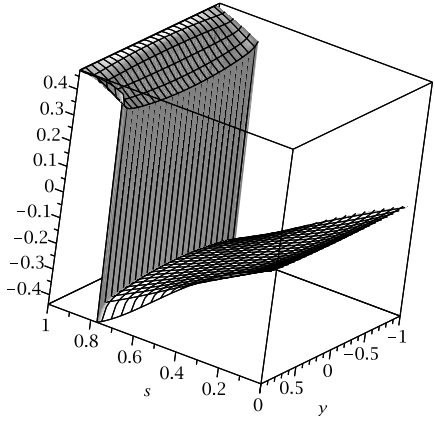} 
	
	\caption{\label{fig4} {\it Plot of  $a_r$, as function of   $s$ and  $y$, with $j=0.1$ ,  $\tau=2.1$.}}
\end{figure}

\begin{figure}[h]
	\includegraphics[scale=0.31]{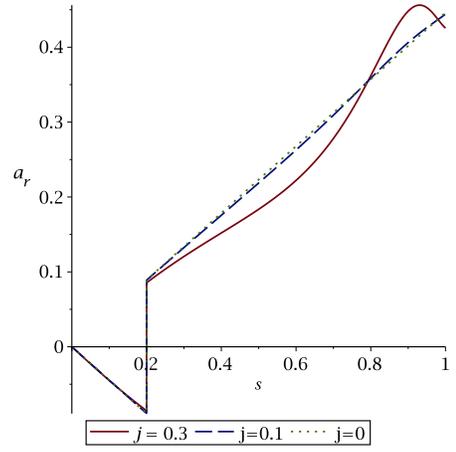} 
	
	\caption{\label{fig5} {\it Curves depicting $a_r$  as function of $s$, for different values of the rotation parameter   $j$,  with $\tau=2.24$,   $y=0.5$.}}
\end{figure}

\begin{figure}[h]
	\includegraphics[scale=0.31]{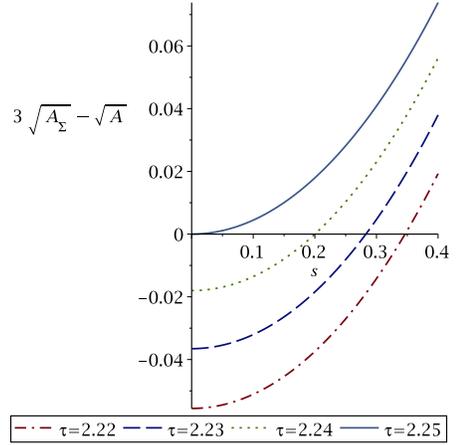} 
	\caption{\label{fig6} {\it   Curves delimiting the range of values of $s$, for which the radial pressure is negative, (the curves cut the axis at $s_c$), for different values of $\tau$. }}
\end{figure}

\begin{table}[htp]
	\caption{$s_{c}$ denotes the value of the dimensionless radial coordinate $s$, enclosing the region where  negative radial pressure appears, this coincides with the region within  which the radial acceleration $a_r$, becomes negative.}
	\begin{center}
		$P<0\ , \ a_r<0$ iff $s \in [0,s_c]$
		
		\begin{tabular}{cc}
			\hline\hline
			$  \tau$ \hspace*{2cm}&$s_c $   \\  \hline
			2.22 \hspace*{2cm}&0.3464  \\  \hline
			2.23 \hspace*{2cm}&0.2828  \\  \hline
			2.24\hspace*{2cm} &0.1999  \\  \hline
			2.25\hspace*{2cm}&0 
			\\ 
			\hline\hline
		\end{tabular}
	\end{center}
	\label{sc}
\end{table}

\newpage

\section{The  fragmentation (cracking) of the source}
In this section we shall tackle a different physical issue, related to our source.

Let us  consider our source as an initial configuration, submitted to perturbations, under which it is a priori unstable. Then we shall evaluate the source  immediately after leaving the equilibrium, where ``immediately'' means at a time scale smaller than the hydrostatic time scale. 

For this purpose, let us first calculate the kinematical variables of the source, and the evolution equation for the  expansion scalar (Raychaudhuri equation).

Since we choose the fluid to be comoving in our coordinates, then we may write the four velocity as:
\begin{equation}
V^\alpha =(\frac{1}{\sqrt{-g_{00}}},0,0,0); \quad  V_\alpha=(-\sqrt{-g_{00}},0,0,\frac{g_{30}}{\sqrt{-g_{00}}}),
\label{m1}
\end{equation}
whereas for the vorticity vector we have:
\begin{equation}
\omega_\alpha=\frac{1}{2}\,\eta_{\alpha\beta\mu\nu}\,V^{\beta;\mu}\,V^\nu=\frac{1}{2}\,\eta_{\alpha\beta\mu\nu}\,\Omega
^{\beta\mu}\,V^\nu,\label{vomega}
\end{equation}
where $\Omega_{\alpha\beta}=V_{[\alpha;\beta]}+a_{[\alpha}
V_{\beta]}$ and $\eta_{\alpha\beta\mu\nu}$ denote the vorticity tensor and the Levi-Civita tensor respectively.

For the covariant derivative of the four--velocity we have the well known expression:

\begin{equation}
V_{\alpha;\beta}=\sigma_{\alpha \beta}+\Omega_{\alpha \beta}-a_\alpha V_\beta+\frac{1}{3}h_{\alpha \beta}\Theta,\label{ps}
\end{equation}
where as usual, $\Theta$, $\sigma_{\alpha \beta}$ and $a_\alpha$ denote the expansion scalar, the shear tensor and the four--acceleration respectivley, and are  defined as:

\begin{eqnarray}
a_\alpha=V^\beta V_{\alpha;\beta},\qquad \Theta=V^\alpha_{;\alpha}.
\label{acc}
\end{eqnarray}

\begin{equation}
\sigma_{\alpha \beta}= V_{(\alpha;\beta)}+a_{(\alpha}
V_{\beta)}-\frac{1}{3}\Theta h_{\alpha \beta}. \label{acc}
\end{equation}
and $h_{\alpha \beta}$ is the projector onto the hypersurface orthogonal to the four--velocity.

Now, the Ricci identities for the vector $V_\alpha$ read
\begin{equation}
R^\mu_{\alpha \beta \nu} V_\mu = V_{\alpha; \beta; \nu} - V_{\alpha; \nu; \beta} ,
\label{Ri}
\end{equation}
then using (\ref{acc}) we obtain

\begin{eqnarray}
&&\frac{1}{2}R^{\rho}_{\alpha \beta \mu} V_\rho = a_{\alpha;[\beta}V_{\mu]} + a_{\alpha}V_{[\mu;\beta]}+ \sigma_{\alpha[\beta;\mu]} + \Omega_{\alpha[\beta;\mu]}\nonumber \\&+&\frac{1}{3} h_{\alpha[\beta}\Theta_{,\mu]} +\frac{1}{3}\Theta h_{\alpha[\beta;\mu]}.
\label{32}
\end{eqnarray}
Contracting Eq.(\ref{32}) with $V^\beta g^{\alpha \mu}$ , we find the Raychaudhuri equation for the evolution of the expansion
\begin{equation}
\Theta_{;\alpha} V^\alpha + \frac{1}{3}\Theta^2 + \sigma^{\alpha \beta} \sigma_{\alpha \beta}-\Omega^{\alpha \beta} \Omega_{\alpha \beta} - a^\alpha_{;\alpha} = - V_\rho V^\beta R^\rho_\beta .
\label{Ra}
\end{equation}

Let us now assume that our source is initially stationary (at some $t=0$), implying that both the shear and the expansion vanish. However, immediately after leaving the equilibrium,  these quantities are still negligible, but no so their time derivatives, since  as we have already mentioned, ``immediately'' means on a time scale which is smaller than the hydrostatic time scale.

Then, at the time scale under consideration, (\ref{Ra}) becomes
\begin{equation}
\Theta_{;\alpha} V^\alpha -\Omega^{\alpha \beta} \Omega_{\alpha \beta} - a^\alpha_{;\alpha} = - V_\rho V^\beta R^\rho_\beta .
\label{Rai}
\end{equation}

In order to evaluate the time derivative of the expansion scalar from (\ref{Rai}), we need first to extract some information from the conditions of the vanishing of the expansion scalar and the shear tensor.

	Thus,  the condition $\Theta=0$ produces
	\begin{equation}
	4 A \dot{\hat g} -6 A \dot{\hat a}-  \dot{A}=0,
	\end{equation}
	where the dot  over the functions denotes time derivative, whereas $\sigma_{\alpha \beta}=0$ implies 
	\begin{eqnarray}
		 A \ \dot{\hat g} -  \dot{A}&=&0 \nonumber\\
		 2 A \  \dot{\hat g}+ \dot{A}&=&0 \nonumber\\
		 4 A \ \dot{\hat g} -  \dot{A}&=&0.
	\end{eqnarray}
	
The above equations  impose the constraints  $ \dot{\hat a}= \dot{\hat g}= \dot{A}=0$ on the metric functions, these conditions will be used  when we evaluate  (\ref{Rai}) as well as the Einstein equations.
	
	Then the time derivative of the expansion scalar immediately after the source leaves the equilibrium leads to
	\begin{equation}
	\dot{\Theta}=\frac{e^{-2\hat a}}{Z^2}\left(-3\ddot{\hat a}-\frac{\ddot{A}}{2A}+2\ddot{\hat g}\right).
	\label{timeTheta}
	\end{equation}
	
	The calculation of the energy-momentum tensor from the Einstein equations for  the system out of equilibrium may be written as the sum of the terms   corresponding to the equilibrium state $T_{\alpha (equi)}^{\beta}$  and  terms  which correspond to the system out of equilibrium ($oeq$), thus we write:
	\begin{equation}
T_{\alpha}^{\beta}=T_{\alpha (eq)}^{\beta}+T_{\alpha (oeq)}^{\beta}.
	\end{equation} 
	The components of these last terms which contain second time  derivatives of  the metric functions (that are not null) read
	\begin{eqnarray}
	T_{0}^{3}(oeq)&=&\frac{\Omega e^{2\hat a}}{8 \pi r^2 \sin^2\theta}\left(-2\ddot{\hat a}-\frac{\ddot{A}}{2A}+2\ddot{\hat g} \right)\nonumber\\
	T_{3}^{3}(oeq)&=&\frac{1}{8 \pi}\left(-\frac{\Omega^2 e^{2\hat a}}{r^2 \sin^2\theta}+\frac{e^{-2\hat a}}{Z^2}\right)\left(-2\ddot{\hat a}-\frac{\ddot{A}}{2A}+2\ddot{\hat g} \right)\nonumber\\
	T_{1}^{1}(oeq)&=&\frac{e^{2\hat a}}{8 \pi r^2 \sin^2\theta}\left(-\Omega^2\ddot{\hat g}-\frac{\Omega^2 \ddot{A}}{Z 4\sqrt{A}}-\Omega \ddot{\Omega}-\dot{\Omega}^2 \right)+\nonumber\\
	&+&\frac{e^{-2\hat a}}{8 \pi Z^2}\left(\ddot{\hat g}-2\ddot{\hat a}  \right)
\label{einsteint}
	\end{eqnarray}
	
	It is easy to see from the equations above  that 
	\begin{equation}
\Omega \ 	T_{0}^{3}(oeq)+	T_{3}^{3}(oeq)=\frac{1}{8 \pi}\dot{\Theta}+\frac{e^{-2\hat a}}{8 \pi Z^2} \ \ddot{\hat a},
\label{raikerr}	
\end{equation}
	and the terms from the stationary case lead to the identity
	\begin{equation}
	\Omega \ 	T_{0 (eq)}^{3}+	T_{3  (eq)}^{3}=\frac{e^{2\hat a-2\hat g}}{8 \pi}\left(8\pi P-\hat{p}_{zz}+\delta J_{+} \right)
	\end{equation}
	with the notation used in \cite{kerrinterior}.
	
	Let us now first consider the spherically symmetric case ($j=0$). In such a case the evolution equation for the expansion scalar,  immediately after leaving the equilibrium, reads 
	\begin{equation}
	\dot{\Theta}=\frac{1}{Z^2}\left(-\frac{\ddot{A}}{2A}\right),
	\label{timeThetaesf}
	\end{equation}
	which, after using the Eisntein equations, becomes
	\begin{equation}
		T_{3}^{3}(oeq)=\frac{1}{8 \pi}\dot{\Theta}.
	\label{ultima}
\end{equation}
	
We see from (\ref{ultima}), that the sign of $\dot \Theta$ for any piece of material, is the sign of $T_{3}^{3}(oeq)$. Now, if we recall that in the spherically symmetric case,  $T_{3}^{3}(eq)=P$, the physical consequence of  (\ref{ultima}) is quite obvious: if the exit from the equilibrium state of any piece of material, is produced by an increase  (decrease) in the pressure, then  that region  will tend to expand (contract). In other words, all fluid elements in that region  will experience  an overall expansion (contraction) once the system leaves the equilibrium, while no  fragmentation (cracking) \cite{cracking} will be observed. 

This absence of cracking in the spherically symmetric limit, with homogeneous energy density distribution, is expected since, it can be rigorously shown that changes in the sign of $\dot \Theta$ are a necessary condition for the occurrence of cracking (see \cite{cr}). Furthermore as has been confirmed in several numerical studies, cracking in spherically symmetric configurations with homogeneous energy density, requires the perturbation of the pressure isotropy \cite{grg}, whereas the spherically symmetric limit of our source is strictly isotropic in the pressure..

Let us now turn back to the  situation under consideration (non--spherical, rotating source). The corresponding equation for the time derivative of the expansion is (\ref{raikerr}). As  is apparent from this equation, now the sign of $\dot \Theta$ not only depends on the variation of one of the diagonal pressure terms, but also on $T_{0}^{3}(oeq)$ and the value (and sign) of the vorticity. This opens the way for a large   number of scenarios, for which the sign of  $\dot \Theta$ changes within the fluid distribution, giving rise to the possibility of the fragmentation (cracking) \cite{cracking} of the source.

\section{Discussion}
We may summarize the results obtained in this work, through the following points:
\begin{itemize}
\item There exists a bound for the maximal value of the surface gravitational potential (minimal value of $\tau$), producing a bound in the maximal value of the surface redshift , that can be observed from the source discussed here.
\item The above mentioned bound was established by detecting the critical value of $\tau$ for which negative acceleration appears, within the fluid distribution. 
\item This critical value, defines exactly the same range of values of the radial coordinate, for which the radial pressure becomes negative. This fact, reinforces  further the physical relevance of the acceleration tensor introduced in \cite{maluf}.
\item The source under consideration, allows the possibility of fragmentation (cracking), once  the equilibrium state has been abandoned. This scenario depends strictly on the non--spherical aspects of the source. In particular it emphasizes the physical relevance of the $T^3_0$ component of the energy--momentum tensor, which as it has been discussed in \cite{kerrinterior}, represents a distinct physical property of rotating fluids.
\end{itemize}
\section{Acknowledgments}
This  work  was partially supported by the Spanish  Ministerio de Ciencia e
Innovaci\'on under Research Projects No.  FIS2015-65140-P (MINECO/FEDER), and the Consejer\'\i a
de
Educaci\'on of the Junta de Castilla y Le\'on under the Research Project Grupo
de Excelencia GR234.

\end{document}